\documentclass[12pt,preprint]{aastex}

\usepackage{graphicx}
\usepackage{subfigure}

\begin{document}

\title{Non-LTE Spectra of Accretion Disks Around Intermediate-Mass Black Holes}

\author
{Yawei Hui, Julian H. Krolik} \affil{Department of Physics and
Astronomy, Johns Hopkins University, Baltimore, MD 21218;
\\ ywhui@pha.jhu.edu, jhk@tarkus.pha.jhu.edu}
\author{Ivan Hubeny}
\affil{Department of Astronomy and Steward Observatory, University
of Arizona, Tucson, AZ 85721; \\ hubeny@aegis.as.arizona.edu}

\begin{abstract}
We have calculated the structures and the emergent spectra of
stationary, geometrically thin accretion disks around 100 and 1000
$M_{\odot}$ black holes in both the Schwarzschild and extreme Kerr
metrics.  Equations of radiative transfer, hydrostatic equilibrium,
energy balance, ionization equilibrium, and statistical
equilibrium are solved simultaneously and consistently.  The six
most astrophysically abundant elements (H, He, C, N, O, and Fe)
are included, as well as energy transfer by Comptonization.  The
observed spectrum as a function of viewing angle is computed
incorporating all general relativistic effects.  We find that, in
contrast with the predictions of the commonly-used multi-color
disk (MCD) model, opacity associated with photoionization of heavy
elements can significantly alter the spectrum near its peak.
These ionization edges can create spectral breaks visible
in the spectra of slowly-spinning black holes viewed from almost
all angles and in the spectra of rapidly-spinning black holes seen
approximately pole-on.  For fixed mass and
accretion rate relative to Eddington, both the black hole spin and
the viewing angle can significantly shift the observed peak energy
of the spectrum, particularly for rapid spin viewed obliquely or
edge-on.  We present a detailed test of the
approximations made in various forms of the MCD model.  Linear
limb-darkening is confirmed to be a reasonable approximation for
the integrated flux, but not for many specific frequencies of interest.
\end{abstract}


\section{Introduction\label{sec:intro}}

Intermediate-mass black holes (IMBHs) are black holes more massive
than the stellar-mass black holes found in galactic binaries
($\sim 10$~$M_{\odot}$) and smaller than the super-massive black
holes found in active galactic nuclei ($\sim
10^6$--$10^9$~$M_{\odot}$).  Whether IMBHs actually exist is at
present an open question.

It has been suggested that IMBHs can be found in Ultra-Luminous
X-ray sources (ULXs)
(\citealt{colbert1999,miller2002,miller2003}).  ULXs are off-center
point-like X-ray objects in nearby galaxies with luminosities
$\sim 10^{39}$-$10^{41}$~erg~s$^{-1}$ in the 2-10~keV band. First
detected in Einstein observations of nearby spiral galaxies (see,
e.g.~\citealt{ft1987}), the list of known examples grew with
\textit{ROSAT}/HRI observations and has been further extended with
\textit{Chandra} and \textit{XMM-Newton} data (see,
e.g.~\citealt{cp2002,swartz2002,stobbart2004,nolan2004}).  Their
variability proves that, whatever their nature, they must be very
compact (\citealt{sch2001}).

If their power is produced by
accretion, and if their X-ray emission is isotropic and
sub-Eddington ($L_{E}=1.3\times 10^{38}(M/M_{\odot}$)~erg~
s$^{-1}$), the mass of the central object must exceed a few tens
to $10^4~M_{\odot}$, depending on the individual object.  On the
other hand, if any of these assumptions are violated, other
explanations for these objects are possible.  Examples of such
models include anisotropic emission (\citealt{king2001}) or
radiation pressure-dominated super-Eddington accretion
(\citealt{begalman2002}), although there at least a few
ULXs with X-ray luminosities too high to be explained in this way
(see, e.g., the review by \citealt{cm2004}).  Possible constraints
on the degree of anisotropy come from observations of nebulae
surrounding ULXs, which are often quite symmetric
(\citealt{pm2003,kaaret2004}).  In this paper, we explore the
possibility that ULXs are, indeed, sub-Eddington accreting
intermediate-mass black holes.

Most ULX X-ray spectra can be described as a sum of two
components.  At low energies (a few hundred eV to several keV), the
spectrum is dominated by an apparently thermal component whose
characteristic temperature is typically $\simeq 0.2$~kev.  On the
other hand, at high energies ($\sim 1$~keV -- 10~keV), the
spectrum is dominated by a power law with photon spectral
index in the range 1.7--2.3 (see,
e.g.~\citealt{miller2003,miller2004,roberts2004}).
It is the goal of this paper to provide the most complete
prediction of the thermal component done to date.

In past work, the thermal component has often been described in
terms of the ``multi-color disk" (MCD) model (see,
e.g.~\citealt{mitsuda1984,makishima1986,shimura1995}, etc.).  In
this model, it is assumed that the total output spectrum of an
accretion disk is a sum of spectra from a series of radial rings.
In this model's simplest form, each ring is supposed to radiate as
a blackbody with an effective temperature, $T_{eff}$, determined
by a thin-disk model.  At large radii, $T_{eff} \propto r^{-3/4}$,
where $r$ is the radius of a particular ring.  In practice, a
variety of variants of this model are employed. Sometimes
$T_{eff}$ is taken to scale as $r^{-3/4}$ all the way to an inner
edge, within which no radiation is produced, and sometimes its
radial scaling is taken from more detailed models (employing
either Newtonian or relativistic physics) that assume a
dissipation cut-off at the innermost stable circular orbit (ISCO).
Sometimes the local ring spectra are supposed to be ``dilute"
blackbodies (see \S 3.1 for a discussion of the basis of this
approximation). Sometimes the intensity distribution is assumed to
be isotropic in the fluid frame as in a true blackbody, and
sometimes it is supposed to follow a guessed limb-darkening law
(\citealt{gierlinski2001}). Sometimes relativistic Doppler shifts
and trajectory-bending are taken into account
(\citealt{gierlinski2001,li2004}), but not always (e.g.
\citealt{makishima2000}). There has been only a small amount of
effort hitherto to {\it predict} rather than {\it assume} the
spectra radiated by individual disk rings.
\citealt{ross1996}, \citealt{merloni2000} and \citealt{fabian2004}
have employed real transfer, but including only Thomson and
free-free opacity (the last paper working specifically in the
context of intermediate-mass black holes); \citealt{davis2004} have
employed the same method as we, but applied to disks around
lower-mass black holes.

In this paper, we report calculations that predict the spectra of
IMBH accretion disks on the basis of complete atmosphere models
for each individual radial ring.  Assuming a smooth vertical
profile, at each radius we solve the
equations of hydrostatic balance, statistical equilibrium,
radiation transfer, and energy conservation in order to arrive at
the emergent intensity in the fluid frame as a function of
frequency and angle.  All ionization stages of C, N, O, and Fe are
included, as well as H and He.  Because we do not allow for the
sort of intense heating required to make one, our models do not
include any hard X-ray-producing corona, even though a significant
hard X-ray component is generally seen in ULX spectra.  Nonetheless,
there can still be mild temperature increases above the photosphere,
and we compute any Comptonization taking place there.  The
absence of a truly hot ($\sim 100$~keV) corona from out model
also means we may be underestimating the photoionization rate
in the disk's upper layers.  A better treatment of disk coronae
and their effects on disk atmospheres must await greater knowledge
of their structure.  In the last step of our calculation,
we apply a transfer function that takes into account all general
relativistic effects to obtain the spectrum seen by
distant observers as a function of angle from the black hole
rotation axis.  In order to show clearly what is new about our
results, we make a detailed comparison between our predictions and
selected variants of the MCD model.

In \S 2 we describe in some detail how we performed our
calculations.  Specific results are presented in \S 3.
We discuss these results in \S 4, with special attention
given to checking the quality of the approximations and
assumptions made in the MCD model.  In \S 5, we summarize.

\section{Description of our Non-LTE disk model}\label{sec:model}

\subsection{Accretion disk model}\label{sec:dmodel}

The physics involved in our calculation of the structure and
radiation field of an accretion disk around an IMBH is very nearly
the same as that described in a series of papers by Hubeny et~al.
(see, e.~g.~\citealt{hubeny1, hubeny2, hubeny3, hubeny4}). The
principal new element in our work is that we apply this model to
accretion disks whose central masses are $10^2$ and $10^3
M_{\odot}$. The earlier work focussed on quasars, and therefore
considered disks with central masses in the range $10^6 - 10^9
M_{\odot}$.  In an effort parallel to this work, \citealt{davis2004}
used the same model to study the output spectra from
disks accreting onto a $10 M_{\odot}$ black hole.

We assume that these disks are axisymmetric, time-steady and
geometrically thin.  The standard $\alpha$-model of accretion disks
(see, e.~g.~\citealt{sunyaev1973,novikov1973}) is adopted, i.e.,
in which the vertically averaged viscous stress, $<t_{r
\phi}>\equiv h^{-1}\int_0^h t_{r \phi}dz$, is related to the
vertically averaged total pressure, $<P>$, by the Shakura-Sunyaev
$\alpha$, $<t_{r \phi}>=\alpha <P>$.  In our calculation, the value
of $\alpha$ is taken as 0.01, which is close to the result coming
from simulations of magneto-rotational instability in unstratified
shearing boxes (see, e.~g.~\citealt{balbus1998}).  Total pressure,
which is composed of the gas pressure and radiation pressure, is
determined by solving the hydrostatic equilibrium equation
neglecting self-gravity and assuming a geometrically thin disk.  We
further assume that the energy dissipation rate, whose vertically-integrated
value varies with radius, is constant per unit mass at any given radius.
By ignoring convection
and conduction, we also assume that energy transport occurs solely
by the vertical flux of radiation. At each ring in our
calculation, the effective temperature is (see,
e.~g.~\citealt{krolik1999})
\begin{equation}\label{eq:teff}
T_{eff}=\left[\frac{3GM\dot{M}}{8\pi\sigma r^3}R_R(r)\right]^{1/4},
\end{equation}
where $\sigma$ is the Stefan-Boltzmann constant and $R_R(r)$
incorporates both the effect of the net angular momentum flux
through the disk and relativistic corrections (see,
e.~g.~\citealt{novikov1973}).

The hardest part of the calculation is solving the radiative
transfer equation simultaneously with other structural equations.
Because we make no {\it ad hoc} assumptions about the the
emissivities and opacities or the source function in the disk, we
have to determine the angle-dependent radiation field and the
emissivities and opacities simultaneously. Because we do not
assume local thermal equilibrium (LTE), we must solve the
statistical equilibrium equations to find the population
densities from which the emissivities and opacities can be
calculated.  However, before starting to solve these equations,
information about the radiation field is needed to construct the
rates at which radiation induces changes in atomic state
populations.  To escape this dilemma, we put all of these
ingredients into a self-consistent model that treats the disk
structure, energy balance, statistical equilibrium and radiation
transfer simultaneously.

Besides those considerations about consistency, two other
physically important features are added to our model. One is the
effect of thermal Comptonization.  We incorporate it into our
calculation by introducing an angle-averaged Compton source
function (see, e.~g.~\citealt{hubeny4}), which is constructed
under the Kompaneets approximation. The other feature is
ground-state continuum opacities of all ionization stages in all
the elements included (H, He, C, N, O, Fe).  We assume solar
abundances, as given in \citealt{anders1989}.

Finally, we specify the parameters of the four disks whose spectra
we compute and give a brief explanation of the flow of
calculation.  All our disks accrete at a rate yielding
$L/L_{E}=0.1$.  Two disks are assumed to lie around
Schwarzschild (non-rotating) black holes, two around maximal Kerr
black holes. By maximal Kerr, we mean $a/M = 0.998$, the limiting
ratio of angular momentum to mass estimated by Thorne (1974). For
each of the black hole spin cases, we calculate the light output
from disks surrounding black holes of two different masses, $100M_{\odot}$
and $1000M_{\odot}$.  To produce the required luminosity relative
to Eddington, the mass accretion rate in the Schwarzschild case is
$3.89\times10^{-9}(M/M_{\odot}) M_{\odot}$~yr$^{-1}$, whereas it
is $6.84\times10^{-10}(M/M_{\odot}) M_{\odot}$~yr$^{-1}$ for the
extreme Kerr example.

To compute the integrated spectrum as a function of angle from the
disk axis, we divide the disks into a series of concentric annuli.
The emergent intensity in the fluid frame is computed for each
annulus using TLUSTY (\citealt{hubeny1,hubeny4}), and these
intensities are then combined into frequency- and angle-dependent
fluxes using a general relativistic transfer code
(\citealt{agol1997}).  The transfer code includes Doppler shifting
and focussing due to the large orbital speeds, gravitational
red-shifts, and bending of geodesics.  The radial grid for the
Schwarzschild case (in terms of gravitational radius $GM/c^2$) is
$r = $ 6.5, 7.0, 7.5, 8.0, 8.5, 9.0, 9.5, 10, 11, 12, 13, 14, 15,
16, 17, 18, 20, 30, 40, 50, 60, 100, 200; for the extreme Kerr
disk it is finer: $r=$ {1.4, 1.5, 1.6, 1.7, 2.0, 2.5, 3.0, 3.5,
4.0, 4.5, 5.0, 5.5, 6.0, 6.5, 7.0, 7.5, 8.0, 8.5, 9.0, 9.5, 10,
11, 12, 13, 14, 15, 16, 17, 18, 19, 20, 25, 30, 35, 40, 45, 50,
60, 70, 80, 90, 100, 120, 140, 160, 180, 200, 250, 300, 350, 400,
450, 500. The inner and outer radii are chosen to be just outside
the marginally stable orbit (on the inside) and far enough away
(on the outside) to ensure that the portion of the spectrum within
a factor of ten in frequency of the peak is fully represented.

\subsection{MCD approximation}\label{sec:mcd}

A significant part of our effort will be contrasting our results
with the standard model used in the literature, the multi-color
disk (MCD) model.  This model was first introduced to explain
observed X-ray spectra by \citealt{mitsuda1984}. The basic idea
behind MCD is to assume that the spectrum of an accretion disk
around a black hole is simply a superposition of individual black
body spectra, each corresponding to a specific radius.
Unfortunately, there is a great deal of confusion in the
literature because a number of variations of this idea have been
used, all under the same name.

Before identifying which version we take as our standard of
comparison, we describe these variations, from simplest to most
complex. In the very simplest version, the factor $R_R(r)$
(eq.(\ref{eq:teff})) is set equal to unity everywhere.  Making no
relativistic corrections, one would then predict an observed
spectrum $\propto \nu^{1/3} \exp{(-h\nu/kT_{\rm max})}$, where
$T_{max}$ is the temperature of the innermost ring
(\citealt{lynden-bell1974}). This model is sometimes elaborated in
any of three directions. First, even in Newtonian physics, the
disk effective temperature does not scale $\propto r^{-3/4}$ all
the way to the ISCO and then dive abruptly to zero; the net
angular momentum flux through the disk leads to a roll-over that
begins somewhat outside the ISCO.  The run of effective
temperature with radius is sometimes adjusted to reflect this.  The
detailed form of this roll-over can be further refined by
considerations of relativity (\citealt{novikov1973,page1974}).
Second, the local spectrum may not be exactly black body.  Many
(perhaps most) people using the MCD model suppose that
Comptonization shifts the peak of the spectrum to higher
frequency.  This shift is described in terms of a ``dilution" or
``hardening" factor $f$ so that the mean intensity at the surface
of the disk is
\begin{equation}\label{eq:dilutebb}
J_\nu = f^{-4} B_\nu(fT_{\rm eff}).
\end{equation}
This factor is generally taken to be a constant 1.7, independent
of circumstance (\citealt{shimura1995,gierlinski2001}).  Third,
some assumption must be made about the angular distribution of the
emergent intensity.  Sometimes it is taken to be exactly isotropic in the
fluid frame, as it would be if the spectrum were precisely thermal.
In other cases (e.g.,
\cite{gierlinski2001}), a specific choice of limb-darkening law is
guessed.  In Newtonian physics, these choices suffice to determine
the angular distribution of the flux.  However, relativistic
effects lead to angle-dependent Doppler shifts, beaming,
gravitational redshifts, and trajectory-bending.  Sometimes these
are ignored, sometimes included via exact treatments.

In the rest of our paper, if not specified explicitly, the MCD
approximation we choose to compare to our results is the one used
by \cite{gierlinski2001}.  They include the general relativistic
corrections to both the disk structure and the photon transfer to
distant observers; a diluted black body (hereafter,
DBB) with $f=1.7$; and a linear limb-darkening law in the fluid
frame
\begin{equation}\label{eq:gierlimbdark}
I_\nu^{\rm ff} (\mu) = 0.42(1 + 2.06\mu)f^{-4}B_\nu \left(f T_{\rm
eff}\right).
\end{equation}

\section{Results of calculation}\label{sec:result}

In this section we report the main results of our calculation.  In
subsection \S \ref{subsec:fluid}, we study the flux in the fluid
frame at selected radii.  Next, we employ the GR transfer function
and predict the spectra as seen at various viewing angles (\S
\ref{subsec:observer}).  Limb-darkening is discussed in greater
detail in \S \ref{subsec:limb}, where we contrast fluid-frame and
observer-frame descriptions. In the last two parts of this
section, we study the effects on observed spectra of varying black
hole spin (\S \ref{subsec:metric}) and mass (\S
\ref{subsec:masseffect}).

\subsection{Spectra in the local fluid frame}\label{subsec:fluid}

We begin by showing how Comptonization and heavy element
photoionization opacity affect the spectrum in the fluid frame. To
isolate the effects of heavy elements, we also present the results
of calculations that include only H and He; these are labelled
``H-He disk", whereas calculations with the full complement of
elements are designated ``Fe disk".  The results shown in
Fig.~\ref{fig:fluidflux} are for an accretion disk around a
maximal Kerr black hole with mass $10^2~M_{\odot}$. The annuli
chosen to plot are $r$=1.6, 6.5, 10, 20, 50, 100~$r_g$ with,
correspondingly, $T_{\rm eff}(r)$=6.2, 2.8, 2.1, 1.3, 0.68, 0.41
in $10^6K$.   Here $r_g \equiv GM/c^2$.  Three different predictions
for the emergent fluxes
from these 6 rings are shown: a true blackbody with $T=T_{\rm
eff}(r)$; the spectrum of an H-He disk; and the spectrum of an
Fe-disk.

\begin{figure}[b]
\centering
\includegraphics[width=1.0\textwidth]{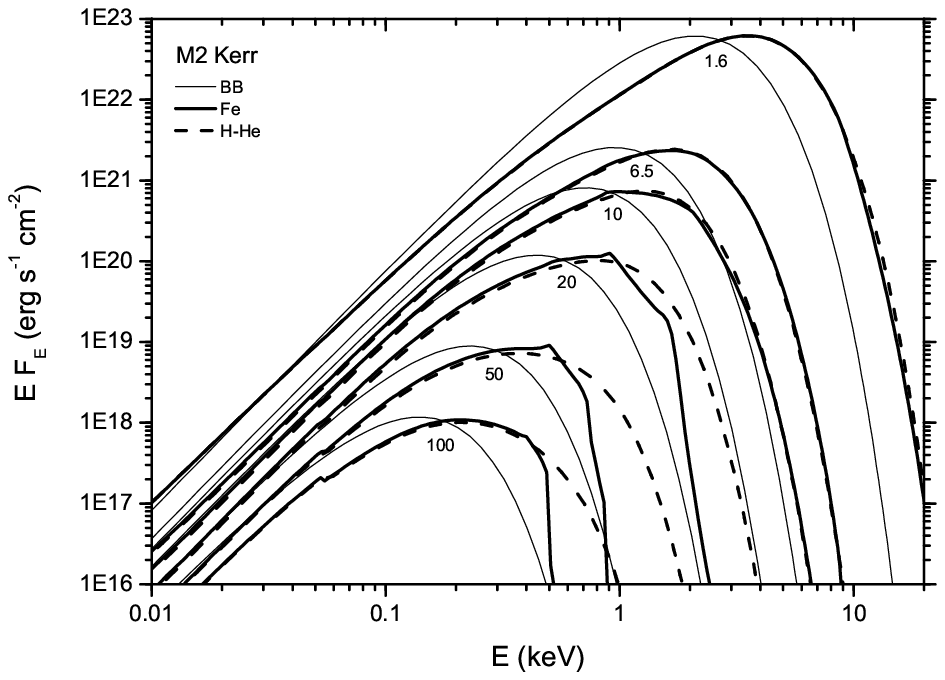}
\setlength{\abovecaptionskip}{0pt}
\setlength{\belowcaptionskip}{10pt} \caption{Local flux of H-He
and Fe disks in the fluid frame. The central mass of the BH is
$10^2M_{\odot}$ and it rotates with spin parameter $a/M
=0.998$. The thin curves represent the blackbody
spectrum with the local effective temperature, while the thick curves
show the full atmosphere predictions.  Two chemical compositions are
displayed, pure H-He atmospheres by dashed curves, solar abundances
by solid ones.  Six annuli are shown, each of them marked by
its radial coordinate in units of $r_g$.}\label{fig:fluidflux}
\end{figure}

The first obvious difference between our NLTE spectra and the
blackbody spectra is the blue-shifting of the peak.  This shift
is largely due to electron scattering blanketing of the
thermalization photosphere, an effect most concisely parameterized
in terms of the photon-destruction probability
$\epsilon_\nu\equiv\kappa_\nu^{abs}/(\kappa_\nu^{abs}+\kappa_\nu^{sc})$.
Here $\kappa_\nu^{abs}$ and $\kappa_\nu^{sc}$ are the absorption
and scattering opacity, respectively.  The surface mean intensity
is then (see, \citealt{rybicki1979} or \citealt{mihalas1978})
\begin{equation}\label{eq:surfmeanintensity}
J_{\nu}(0)\approx\sqrt{\langle\epsilon_{\nu}\rangle}~B_{\nu}(T^\ast),
\end{equation}
where $T^\ast$ is the temperature at the thermalization depth.
The local photon-destruction probability in general depends on both
density and temperature, and therefore requires an appropriate
average for use in equation~\ref{eq:surfmeanintensity}.
When $\langle\epsilon_\nu\rangle \ll 1$, the spectrum is thermalized at an
optical depth $\tau^{*}_\nu \simeq (3\langle\epsilon_{\nu}\rangle)^{-1/2}$.
The physical effect underlying equation~\ref{eq:surfmeanintensity} is
that a large ratio of scattering to absorption opacity leads to a thick
scattering layer between the thermalization photosphere and the
surface, suppressing the output.  When this occurs in an isothermal
atmosphere, the resulting spectrum is called a ``modified blackbody".

In a H-He disk, absorption is entirely due to the free-free process,
and so declines monotonically with increasing frequency. The
photon-destruction probability
therefore also declines monotonically with increasing frequency, depressing
the output below Planckian by larger and larger amounts.  However,
in these atmospheres, the temperature increases inward.  The decrease
in $\epsilon_\nu$ at high frequencies therefore {\it raises} $T^\ast$,
which more than compensates for the increased thickness of the
scattering blanket.

\begin{figure}[b]
\centering
\includegraphics[width=1.0\textwidth]{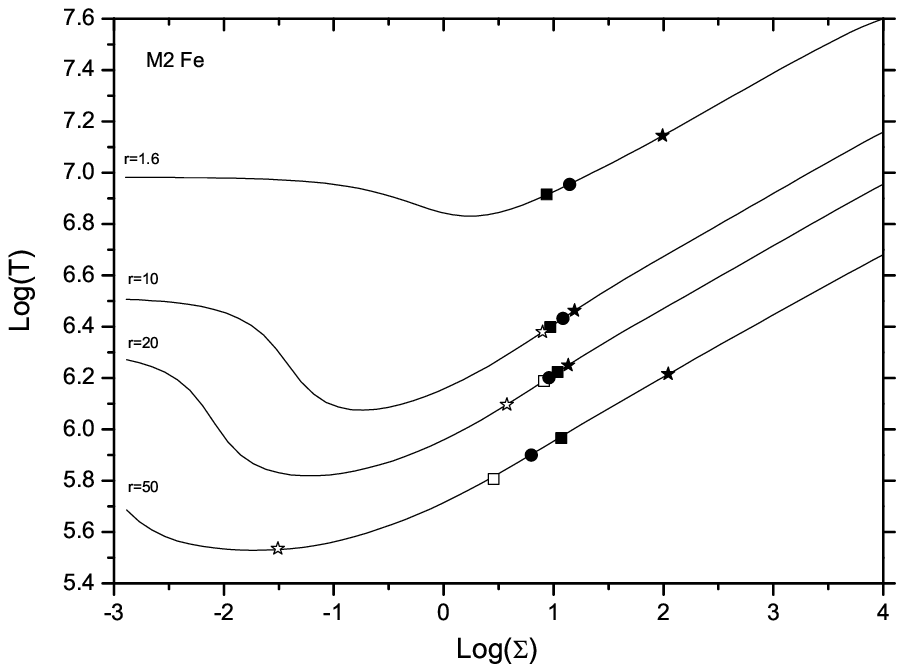}
\setlength{\abovecaptionskip}{0pt}
\setlength{\belowcaptionskip}{10pt}
\caption{Temperature profiles
of four disk annuli ($r=$ 1.6, 10, 20, 50 $r_g$) are shown
together with the positions of the thermalization depths for
energies straddling the ionization edges of CVI (squares) and
OVIII (stars).  Depth into the disk is measured in terms of the
column mass density, $\Sigma\equiv \int_{z}^{\infty} \, dz \, \rho$.
Solid symbols
represent energies just below the ionization edge of the given
element, and hollow symbols just above the ionization edge. The
Rosseland mean thermalization depth is shown by the solid circle
at each radius. There is little temperature contrast between the
thermalization depths immediately above and below the ionization
edges in the innermost part of the disk, but at larger radii the
temperature contrast grows considerably.
 }\label{fig:m2feTauEff}
\end{figure}

Adding heavy elements does little at low frequencies because they
have little opacity there. However, photoionization edges due to
CVI, OVIII and FeXIX (H-like C and O, O-like Fe) appear at 490,
871~eV and 1.53~keV, respectively. These energies
happen to lie near the peak of the emergent spectra for our mass
range of interest (100--$1000M_{\odot}$).  By sharply {\it
increasing} $\epsilon_\nu$ at these energies and above, $T^\ast$
decreases, dimming $J_\nu(0)$. In the inner rings the temperature
is so high that C and O are almost entirely stripped, and this
effect is weak.  Farther out in the disk ($r/r_g \gtrsim 10$),
more H-like ions are
present, and this effect is quite strong.  The net result is a
sharp drop in the high-frequency end of the spectrum emitted by
these outer rings. In Fig.~\ref{fig:m2feTauEff}, we place the
thermalization depths for the CVI and OVIII edges relative to the
the temperature profiles. As the temperature falls toward larger
radii, the temperature difference between the thermalization
depths just below and just above the ionization edges grows.  The
result is an increasingly strong contrast in emergent flux across
the edge.

Interestingly, by comparing eq.~\ref{eq:surfmeanintensity} with
eq.~\ref{eq:dilutebb}, we see the factor
$\langle\epsilon_{\nu}\rangle^{1/2}$ provides an estimate of the
hardening factor used in the formulae of the dilute blackbody
approximation: $f \simeq
1.78(\langle\epsilon_{\nu}\rangle/0.01)^{-1/8}$. The usual choice
for $f$ ($\simeq 1.6$ or 1.7) is therefore approximately
consistent with the physics of local modified blackbody spectra.
It is important to note, however, that $\langle
\epsilon_{\nu}\rangle$ is in general a function of frequency,
whereas $f$ is generally taken to be a constant. In addition, the
DBB model also uses $f$ to estimate the blue shift of the spectral
peak, an effect due to a combination of the internal temperature
gradient and the depth of the thermalization surface.  This second
use is clearly much less well-justified than the first.

Comptonization can also cause the photon energy to be shifted to a
higher value when the temperature is sufficiently high. The
strength of this effect depends primarily on the Compton-$y$
parameter, which can be defined in our context (see,
\citealt{rybicki1979}) as
\begin{equation}\label{eq:comptony}
y=4\Theta~\mathrm{max}[\tau^{*}_\nu,~(\tau^{*}_\nu)^2],
\end{equation}
where $\Theta \equiv kT/(m_e c^2)$ (\citealt{hubeny4}.)
Only for $1.6 \leq r/r_g \leq 10$ and frequencies just below
the ionization edges (maximizing $\tau^*_\nu$) does $y$
exceed unity, and even then not by very much.  Comptonization
therefore seems unlikely to play a large role.  This expectation
is vindicated by the fact that the magnitude of the shift between
the blackbody peak and the peak of the emergent spectrum hardly
changes as a function of radius, although Comptonization should
disappear entirely at large radius.

\subsection{Spectra in the observers' frame at infinity}\label{subsec:observer}

After a GR transformation of the emergent radiation from the local
fluid frame to a distant observer's frame and an integration over
the radius grid, we obtain the spectrum as seen at any particular
angle.  For an overview of the observed spectrum, we display the
angle-averaged observer's frame specific luminosity in
Fig.~\ref{fig:angave}, as predicted by a local blackbody model, a
pure H-He disk, and our solar abundance disk. The atmosphere model
predictions, with or without heavy elements, exhibit a peak in the
range 1--2.5~keV that is significantly blue-shifted with respect to
the spectrum predicted assuming local blackbody emission.  This blue-shift
reflects, of course, the blue-shift already seen in the fluid
frame spectra. As we will see later, different viewing directions
entail varying Doppler shifts, and these can dramatically shift
the location of the peak as seen by any one observer.

However, a smooth peak is not the only feature of this spectrum.
Two atomic features can also be identified, ionization edges of
CVI and OVIII.  Although they appear at 490~eV and 871~eV,
respectively, in the fluid frame, they are stretched and smoothed
by differential Doppler shifts in the angle-integrated spectrum.
The net effect of these edges is to transfer flux from frequencies
above the edges (where the opacity is higher) to frequencies below
(where the opacity is lower).  As a result, the integrated luminosity
from a solar-abundance disk is slightly higher than that of a pure
H-He disk below the OVIII edge and slightly lower above.

\begin{figure}[b]
\centering
\includegraphics[width=1.0\textwidth]{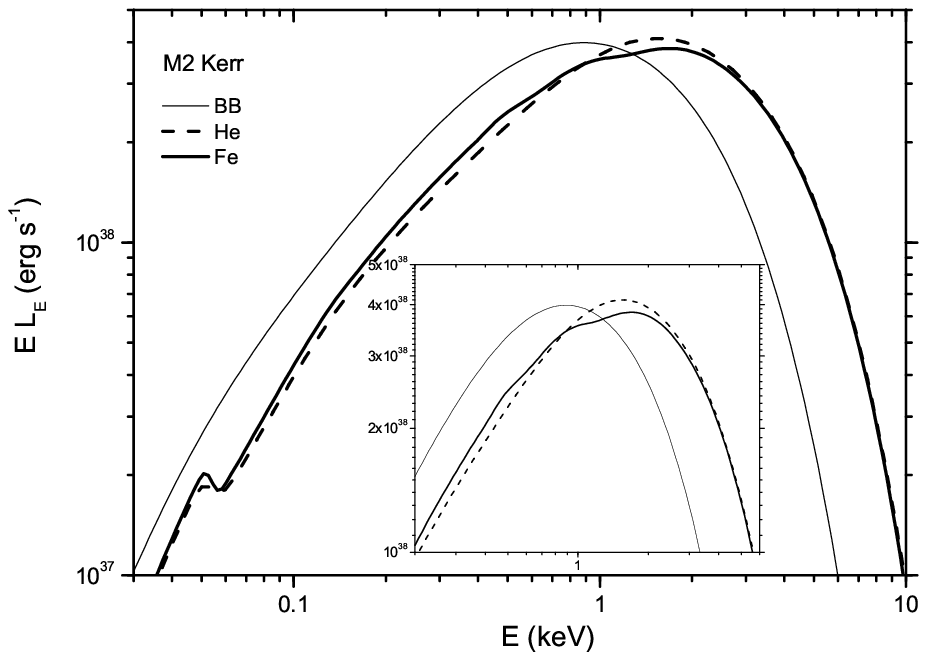}
\setlength{\abovecaptionskip}{0pt}
\setlength{\belowcaptionskip}{10pt} \caption{Angle-integrated
spectra for the H-He (thick dashed) and Fe (thick solid) disks in
the observer's frame.  The central mass of the BH is
$10^2M_{\odot}$ and $a/M =0.998$.  For comparison, the integrated
blackbody spectrum (thin curve) is also shown.  The
inset figure shows the peak area of the same
spectrum in greater detail.}\label{fig:angave}
\end{figure}

\subsection{Angular radiation pattern}\label{subsec:limb}

In this section, we discuss the radiation anisotropy found by our
calculation in both the fluid and the observer's frame. In each
sub-figure of Fig.~\ref{fig:limbdark1.6} and
Fig.~\ref{fig:limbdark}, we plot $I_{\mu}/I_{max}$, i.~e., the
specific intensity normalized by the maximum with respect to angle
at that energy.  The energies whose emergent intensities we choose
here cover the range from $E=0.1$~keV to $E=10$~keV.  Two,
$E=0.49$~keV and $E=0.87$~keV, were selected because they are the
ionization edges of CVI and OVIII, respectively, but we found that
the angle-dependence of the intensity does not depend on whether
the energy chosen is above or below the exact edge energy.

\begin{figure}[ht]
\centering \subfigure[Limb-darkening in the fluid
frame at $r=1.6r_g$.]{ \label{fig:limbdark1.6:a}
\includegraphics[width=0.6\textwidth]{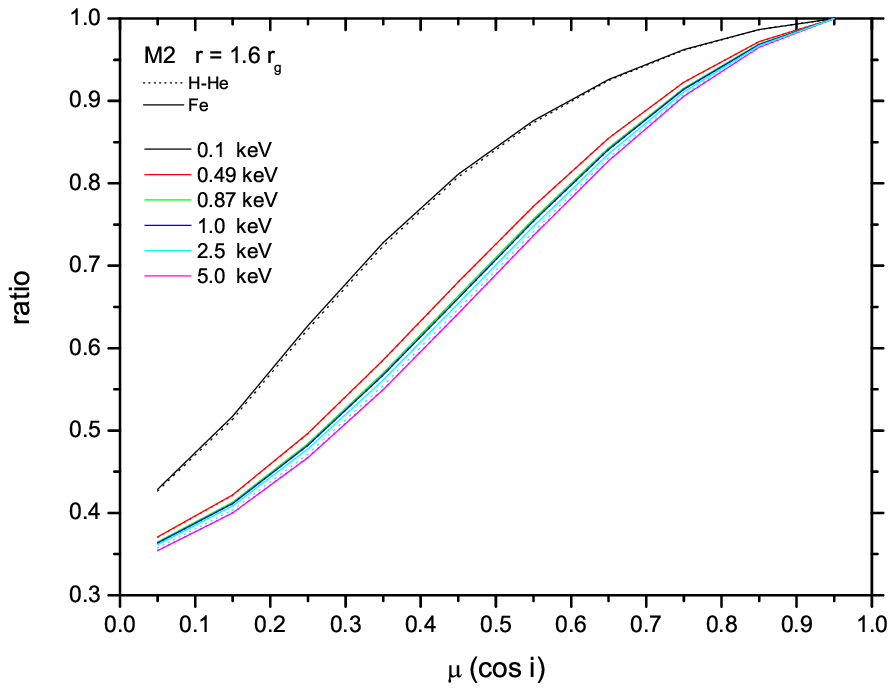}}
\hspace{1in} \subfigure[Similar presentation as in part (a) except
at $r=100r_g$.]{ \label{fig:limbdark1.6:b}
\includegraphics[width=0.6\textwidth]{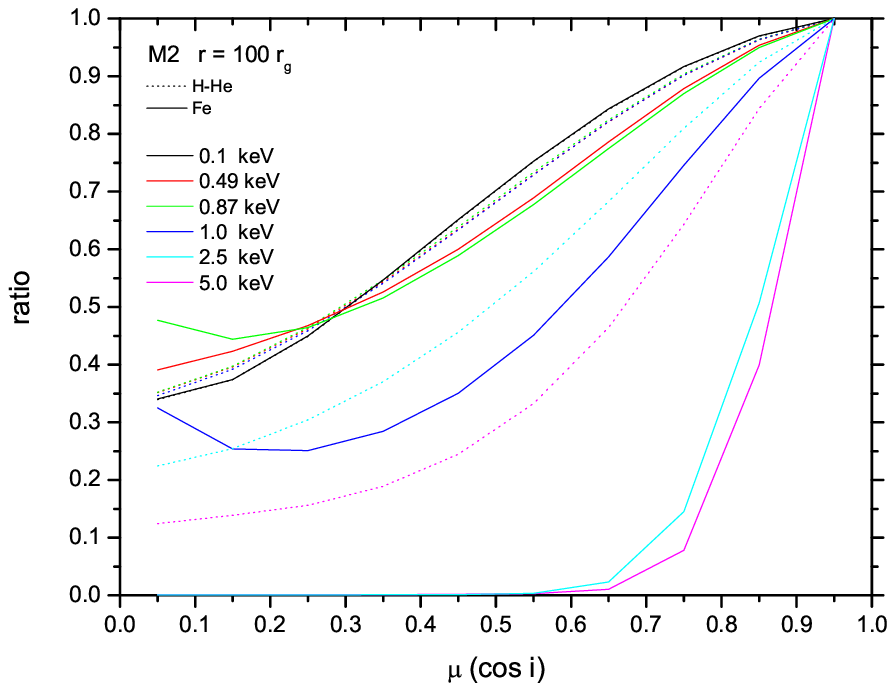}}
\caption{Limb-darkening behavior in the local fluid frame for an
accretion disk with central BH of $10^2~M_{\odot}$ and maximum
spin. Curves in different colors represent different emission
energies. Solid and dotted curves are for the Fe and H-He disks,
respectively.  Each curve is normalized separately so that in all
cases the maximum relative intensity is unity.}
\label{fig:limbdark1.6}
\end{figure}

For the most part, the chemical composition of the disk doesn't
dramatically affect the angle-dependence of the intensity in the
fluid frame no matter which case we are talking about.  Roughly
speaking, the intensity obeys a linear limb-darkening law in the
fluid frame for most energies at all radii. The origin of this
linear limb-darkening is very conventional: at larger viewing
angles, the apparent photosphere lies nearer the surface, where
the temperature is lower.  The variation is smooth enough that
$T_{\rm apparent}(\mu)$ can be approximated as linear.  For
energies below the peak of the Planckian (i.e., in the
Rayleigh-Jeans regime), $I_\nu(0) \propto T_{\rm apparent}$,
whence the linear dependence of emergent intensity on $\mu$.

The exception to this rule comes at high energies and large radii.
Beyond 1~keV, $T_{\rm apparent}$ continues to be a linear function
of $\mu$, but the energies now lie on the Wien side of the
Planckian.  In that regime, the dependence of intensity on
temperature is strongly nonlinear. As the energy increases, this
dependence becomes stronger and stronger so that the
limb-darkening law becomes steeper and steeper, as seen in
Fig.~\ref{fig:limbdark1.6:b}.

Fig.~\ref{fig:limbdark} shows the ``limb-darkening" behavior for
several specific energies as seen in the observer's frame.  Quotes
on this term are appropriate because for much of the interesting
energy range the limb, or an oblique angle, may actually be the
point of view where the light is brightest.  Although low energies
do show fairly conventional linear limb-darkening, higher energies
($E \geq 1$~keV) exhibit very different behavior.  At these
energies, there is relatively little pole-to-limb contrast;
indeed, the intensity does not even vary monotonically as a
function of viewing angle.  As the energy increases, the brightest
direction moves from $\mu = 0.8$ (at 1~keV) to $\mu = 0.6$ (at 2.5~keV)
and to $\mu = 0.45$ (at 5~keV).

A competition between two opposing effects leads to this unusual
result. On the one hand, the emergent intensity in the fluid frame
is limb-darkened for the reasons just discussed.  On the other
hand, because orbital motion is in the plane, relativistic beaming
tries to push flux toward the limb.  The flux at high energy is
generated predominantly in the innermost part of disk, so the
importance of relativistic limb-brightening grows as the energy
increases. This competition accounts for all the effects described
in the previous paragraph.

\begin{figure}[t]
\centering
\includegraphics[width=1.0\textwidth]{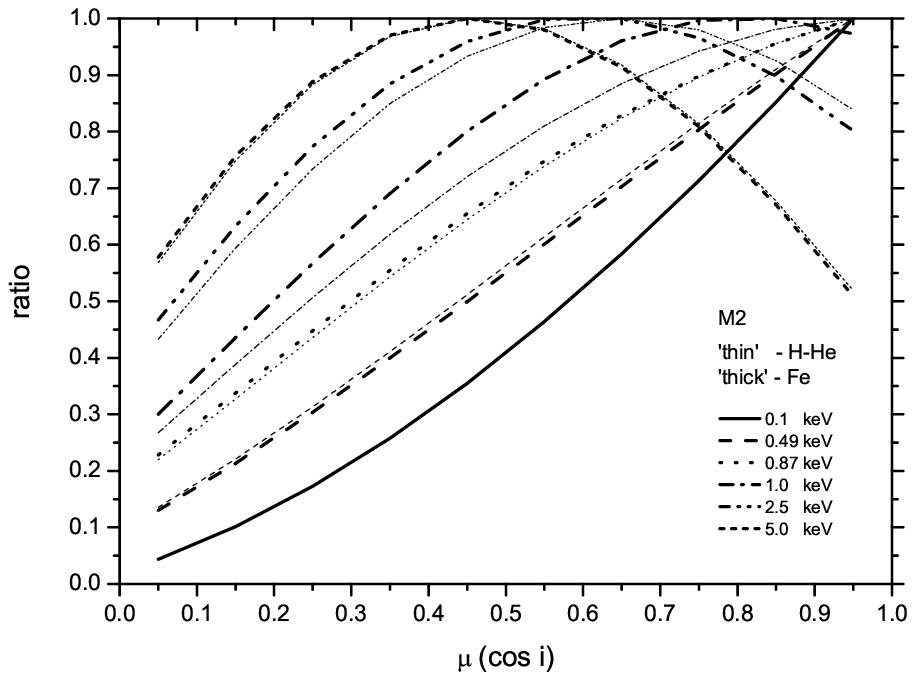}
\setlength{\abovecaptionskip}{0pt}
\setlength{\belowcaptionskip}{10pt} \caption{Limb-darkening
calculated in the observer's frame for an accretion disk
with central BH of $10^2~M_{\odot}$ and maximum spin. Thick and
thin curves are the Fe and H-He
disks, respectively.  Curves in different line types represent
different emission energies.  As in Fig.~\ref{fig:limbdark1.6},
each curve is normalized so that its maximum is unity.}\label{fig:limbdark}
\end{figure}

\subsection{Spectra in different metrics}\label{subsec:metric}

In the previous subsections, we have presented results only from
our calculations of a disk in a maximal Kerr metric.  Not all
black holes spin rapidly, so we have also calculated the spectrum
of a disk around a non-spinning $100 M_{\odot}$ black hole.  The
only change in technique was the use of a coarser radius grid, as
mentioned in section \ref{sec:dmodel}.  In Fig.~\ref{fig:mass100}
we show the results and compare them to the rapidly-spinning case.

\begin{figure}[ht]
\centering
\includegraphics[width=1.0\textwidth]{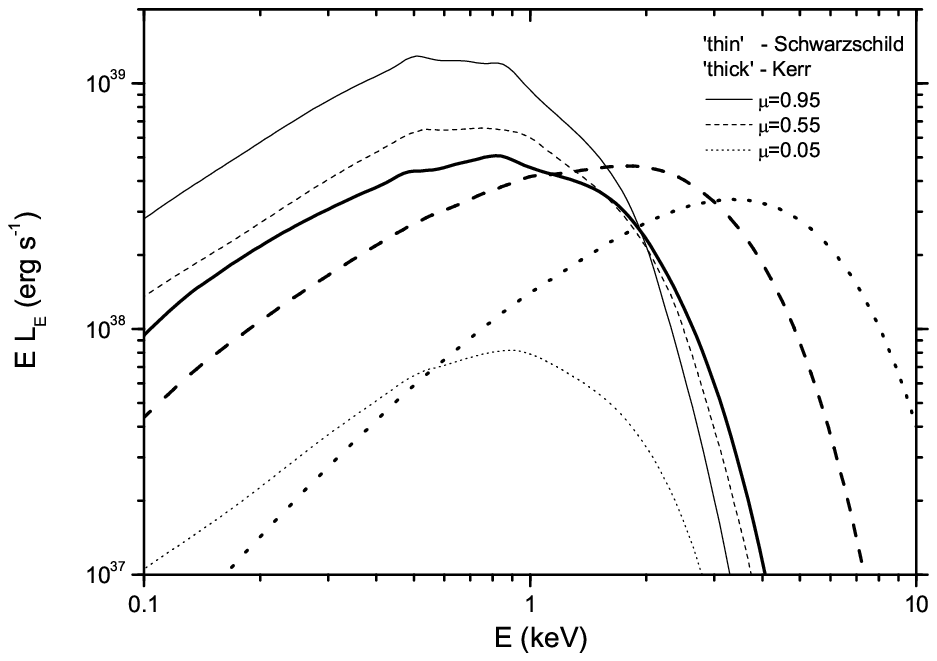}
\setlength{\abovecaptionskip}{0pt}
\setlength{\belowcaptionskip}{10pt} \caption{Spectra of
Fe disks in the observer's frame with central BH mass
$10^2~M_{\odot}$.  The solid, dashed and
dotted line types represent different viewing angles, $\mu=0.95,
0.55, 0.05$, respectively.  Schwarzschild and Kerr metrics
are represented by the thin and
thick curves, respectively.}\label{fig:mass100}
\end{figure}

The biggest contrast between the two cases is that the spectra
are ``cooler" when the spin is slow.  This is a natural outcome
of the fact that in the Schwarzschild metric the innermost stable
circular orbit---in our model, the inner edge of the
disk---is not nearly as deep in the potential as it is in the
maximal Kerr metric.

This contrast in the position of the disk inner edge also leads to
another contrast between the two cases visible in
Fig.~\ref{fig:mass100}: slow spin also leads to much weaker
Doppler shifting, and therefore much weaker dependence of the
observed peak energy on the observer's viewing angle.  The peak
energy for the spectrum emitted by a disk around a Schwarzschild
black hole changes very little as a function of viewing angle,
whereas in the maximal Kerr case, the energy of the peak moves
from 0.81~keV to 3.3keV when $\mu$ changes from 0.95 to 0.05.

What does change with viewing angle in the Schwarzschild case is
the flux itself.  From the pole-on to the edge-on view, the flux
falls by about a factor of 15.  The corresponding ratio for the
Kerr case is only about 1.5. Integrated over
solid angle, the observed bolometric luminosities at infinity are
76\% and 97\% of the assumed $0.1L_{Edd}$ for the Kerr and
Schwarzschild cases, respectively; the missing radiation is in
photons that either return to the disk or are captured by the
black hole (cf. \citealt{agol2000} who computed the capture and
returning radiation fractions
for emission isotropic in the fluid frame).

  In addition, when the relativistic Doppler shifts are comparatively
weak, remnants of the photoionization features imprinted on the fluid
frame spectra survive in the integrated spectrum.  When the black hole
spins slowly, this is the case at all angles, but when the extreme
Kerr metric applies, this is true only for nearly face-on viewing angles.

  The form these features take is a ``platform" around the peak area.
Because it is based on the CVI and OVIII edges, it runs from
$\simeq 500$~eV to $\simeq 870$~eV, with these boundaries changed
only slightly by angle-dependent Doppler shifts. We stress that
the spectral shape in this region is {\it not} a classic
ionization edge; rather, there are breaks in the spectral slope at
these energies and the spectrum in between has relatively little
curvature.

\subsection{Effects of varying central black hole mass}\label{subsec:masseffect}

We have also investigated the effects caused by changing the
central mass of the black hole from $10^2$ (model M2) to
$10^3~M_{\odot}$ (model M3) in the maximal Kerr metric.  The most obvious
contrast is trivial: at fixed $L/L_E$, the luminosity of the disk
around the $1000 M_{\odot}$ black hole is ten times as great as
that of the disk around the $100M_{\odot}$ black hole.

In many respects the shapes of the two spectra are quite similar.
They have similar curvature, and they both have the shelf between
two ionization features when viewed face-on
(Fig.~\ref{fig:mass100_1000}). The principal difference in
spectral shape between the two cases is the downward shift in
energy of the peak as the central mass increases.  The ratio
between the M2 and M3 peak energies varies hardly at all as
a function of viewing angle, as it is $\simeq 1.6$
at $\mu = 0.05$, $\simeq 1.8$ at $\mu = 0.55$, and $\simeq 1.7$
at $\mu = 0.95$. The reason for the
shift, of course, is the well-known scaling law for the central
temperature of an accretion disk around a black hole: $T \propto
M^{-1/4}(L/L_E)^{1/4}$.  In this case, with a factor of 10 between
the two masses, the contrast in central temperature is a factor of
1.8.  Angle-dependent Doppler shifts and, to a lesser degree,
differential atomic opacity effects, cause the angle-dependent
ratio in the peak energy to differ slightly from this ratio.

\begin{figure}[ht]
\centering
\includegraphics[width=1.0\textwidth]{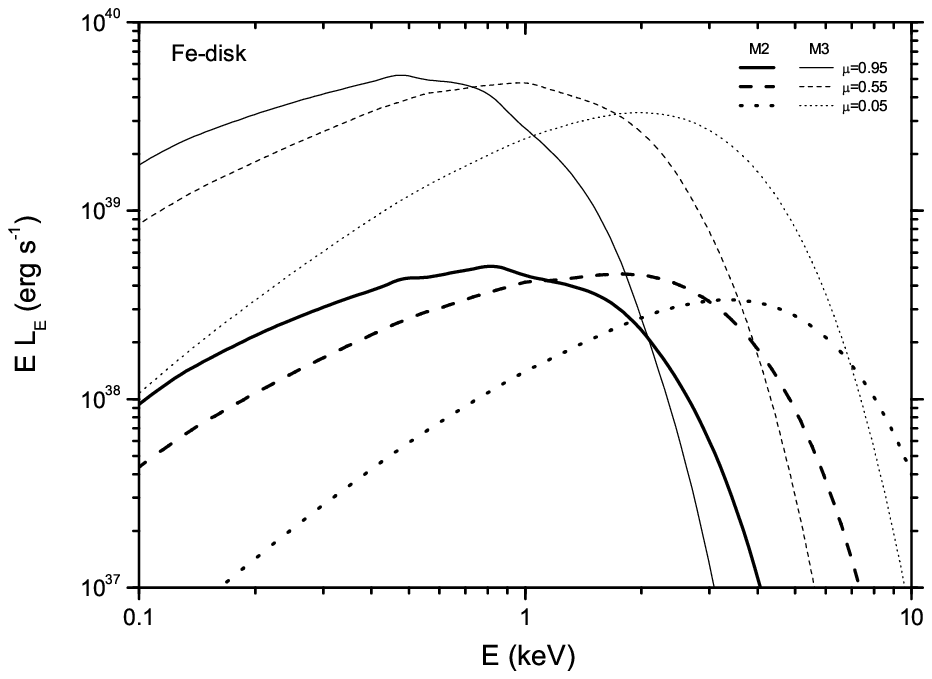}
\setlength{\abovecaptionskip}{0pt}
\setlength{\belowcaptionskip}{10pt} \caption{Spectra of
Fe disks with central BH masses $10^2~M_{\odot}$ and
$10^3~M_{\odot}$ in the extreme Kerr metric. The solid, dashed and
dotted line types represent viewing angles $\mu=0.95,
0.55, 0.05$.  The thick and thin curves are for
$10^2~M_{\odot}$ and $10^3~M_{\odot}$ black holes.}
\label{fig:mass100_1000}
\end{figure}

  There is also a small change in the photoionization features
correlated with the temperature contrast between the two models
with different central masses.  We note that these features are
visible only when the viewing angle is nearly pole-on because
these calculations assume rapidly-spinning black holes.  The CVI
edge is near the spectral peak for the $1000 M_{\odot}$ black
hole, and its associated spectral break is correspondingly more
prominent than the one associated with the OVIII edge.  In the
hotter disk around the $100 M_{\odot}$ black hole, the situation
is exactly reversed: The OVIII edge is nearer the peak and creates
a sharper spectral break.

\section{Discussion}\label{sec:discussion}

\subsection{How do our spectra compare to the MCD model?}

As we have remarked earlier, there are two principal physical
assumptions on which the MCD model rests:  1) The mean intensity
in the fluid frame is a dilute blackbody with a fixed dilution
factor, and 2) the angle-dependence of the intensity in that frame
follows a single predetermined limb-darkening law that is independent
of photon frequency.  In this
section we will assess the quality of those assumptions.

In Fig.~\ref{fig:fluidspec}, we show the emergent flux in the
fluid frame at a number of radii, contrasting our results with
those predicted by a dilute blackbody.  The dilution factor we use
is 1.7, the number chosen by many authors including
\citealt{gierlinski2001}. Without heavy elements, the DBB model is
a reasonable approximation to the actual solution (see,
\citealt{fabian2004}), but with a distinct shift of flux from low
frequencies to high (Fig.~\ref{fig:fluidspec:hhe}). Generally
speaking, the DBB model falls a factor of 2 or so below the
fluid-frame flux computed in the full atmosphere model at
frequencies 10 times below the peak.  It agrees reasonably well
with the full calculation for frequencies within a factor of 3
above or below the peak at large radii ($r \geq 10r_g$), but
predicts $\sim 10\%$ too much flux near the peak at small radii.
At energies well above the peak, the DBB model predicts too little
flux at large radii and too much at small.

These deviations of the DBB from the full calculation are most
likely related to the fact (pointed out
in \S~\ref{subsec:fluid}) that a modified blackbody spectrum is
a better approximation to the emergent intensity in
the fluid frame than a dilute blackbody.  The DBB assumes that the
factor multiplying the Planckian is a constant, but $\sqrt{\epsilon_\nu}
\propto \nu^{-1}$ when the free-free process is the
dominant absorptive opacity.  Although this frequency-dependence is
partially offset by the fact that $T^*$ decreases toward lower
frequency, there is still a net rise toward low frequencies
relative to the constant multiplicative factor of the DBB model.

In addition, smaller deviations can be seen around the peak area,
particularly at larger radius, even in the absence of heavy elements.
For example, the DBB flux near the spectral peak at $r=20r_g$ is 17\%
too high.  We expect that this discrepancy is mostly due to the
other place where $f$ appears, where it is used as a ``hardening
factor" to shift the nominal temperature of the thermal spectrum.
Even in a pure H-He atmosphere, because the temperature gradient
changes from place to place, just as no single dilution factor works
for all radii, neither does any single value of $T^*/T_{\rm eff}$.

\begin{figure}[ht]
\centering \subfigure[Pure H-He atmosphere.]{ \label{fig:fluidspec:hhe}
\includegraphics[width=0.6\textwidth]{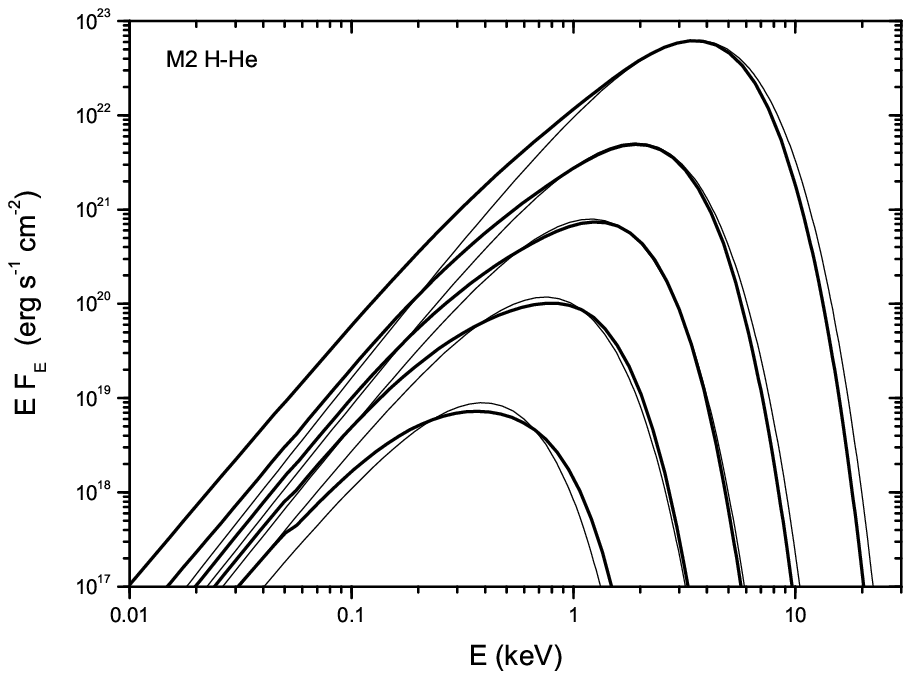}}
\hspace{1in} \subfigure[Atmosphere with H, He, C, N, O, and Fe.]{
\label{fig:fluidspec:fe}
\includegraphics[width=0.6\textwidth]{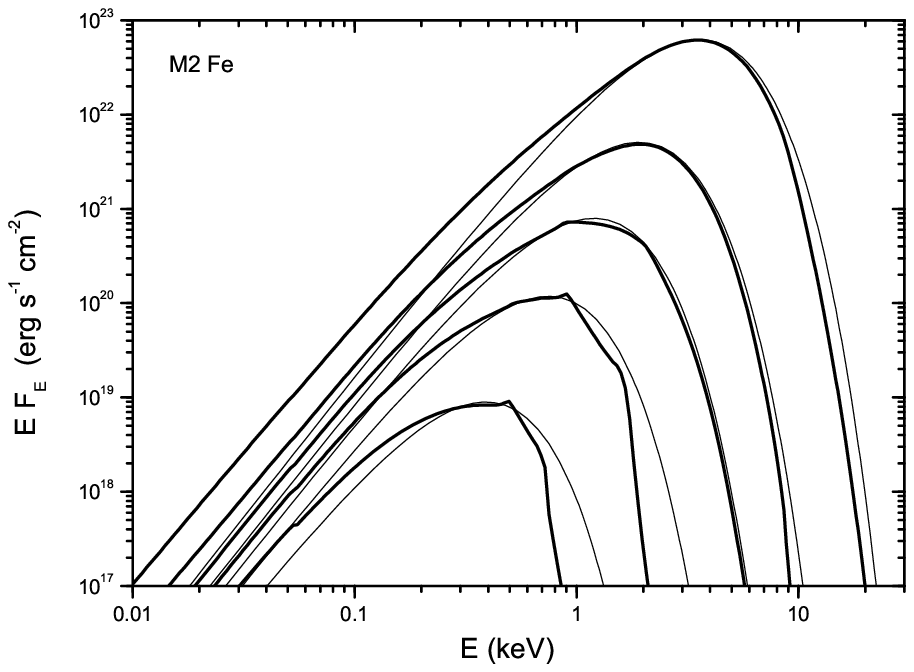}}
\caption{Emergent flux in the local fluid frame for a series
of annuli in an accretion disk around a $10^2~M_{\odot}$ extreme
Kerr black hole. Thick curves are results of our calculations and
thin curves represent the DBB model with the same effective
temperatures. From top to bottom, the radii corresponding to each
pair of curves are: $r/r_g$=1.6, 5.0, 10, 20, 50. }
\label{fig:fluidspec}
\end{figure}

When heavy elements are introduced, the quality of fit provided by
the DBB model worsens sharply, as shown in
Fig.~\ref{fig:fluidspec:fe}.  For frequencies at the peak or below,
the level of agreement is about the same as for the H-He model.
However, in the outer rings, where the atomic photoionization
features become important (cf. \S~\ref{subsec:fluid}), the DBB
model predicts far too much flux at frequencies above the peak.
It is obvious that this sort of discrepancy in spectral shape is
far beyond the ability of a mere change in dilution factor to fix.

We next examine how these local effects carry over into the
observed spectrum.  In Fig.~\ref{fig:face-on}, we present the MCD
spectrum (in the definition of \citealt{gierlinski2001}), and our
two full atmosphere spectra (with and without heavy elements) as
they would be seen face-on, with the frequency range restricted to
the peak region.  We see now the MCD spectrum is approximately
correct only between 0.6 and 0.8~keV and above 1.6~keV, as it
over-predicts the flux from 0.8 to 1.6~keV and
falls too steeply toward lower energies below 0.6~keV.

Just as in the fluid frame, the biggest contrasts between the MCD
spectrum and the full atmosphere are the result of heavy element
opacity.  Because the Doppler effect is reduced to a minimum in
this face-on view, the heavy element features appear with
relatively little smearing, despite the various relativistic
effects.  The MCD model overpredicts the flux near the peak by
about 13\%, but more glaringly, drastically
misses the shape of the spectrum.  Rather than the smoothly
curving shape predicted by the MCD between 0.5 and 2~keV, a full
atmosphere calculation predicts a spectrum with sharp changes in
slope near 490~eV, 560~eV, 810~eV, and 1.5~keV.  Gentler changes
in spectral slope can be found near 0.4, 1.0, and 1.5~keV.
Several of these are clearly associated
with photoionization edges of H-like ions: CVI is at 490~eV, OVIII
at 871~eV, Fe XIX at 1.538~keV. Others likely represent the
end-points of the edge features. Of these, the greatest curvature
is associated with the feature that is strongest in the fluid
frame, the OVIII edge.

\begin{figure}[ht]
\centering
\includegraphics[width=1.0\textwidth]{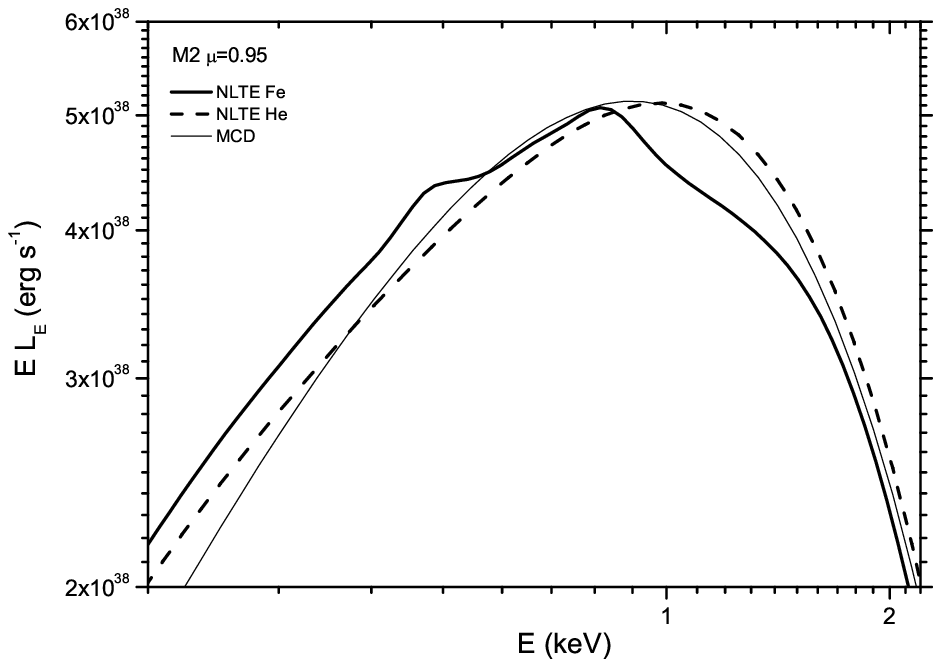}
\setlength{\abovecaptionskip}{0pt}
\setlength{\belowcaptionskip}{10pt} \caption{Spectra of accretion
disks with different chemical compositions in the observer's frame
at infinity. The central BH mass is $10^2~M_{\odot}$ and it is in
the extreme Kerr metric.  Thick curves are results from our
calculations (solid curve is the Fe disk and the dashed is H-He);
the thin curve shows the MCD model modified by
\citealt{gierlinski2001}.  The viewing angle is nearly face-on
($\mu = 0.95$).  Heavy element features create numerous sharp
bends that do not appear when the atmosphere is pure H-He or the
MCD approximation is used.}\label{fig:face-on}
\end{figure}

The face-on view is, of course, the angle at which the
photoionization features are most prominent.  Perhaps a fairer
measure of the typical situation is provided by the spectrum at
$\mu = 0.55$ ($56^{\circ}$ off axis), a direction we find gives
the closest approximation to the flux as integrated over solid
angle.  Fig.~\ref{fig:oblique} shows the spectra predicted by the
same three models as in Fig.~\ref{fig:face-on}, but for
$\mu=0.55$. The contrast between our full atmosphere calculation
with heavy elements and the MCD approximation is now weaker, as
the features are smeared by the strong Doppler broadening made
possible by the Kerr metric. However, even in this case, there is
still a significant spectral inflection between 1 and 2~keV.

\begin{figure}[ht]
\centering
\includegraphics[width=1.0\textwidth]{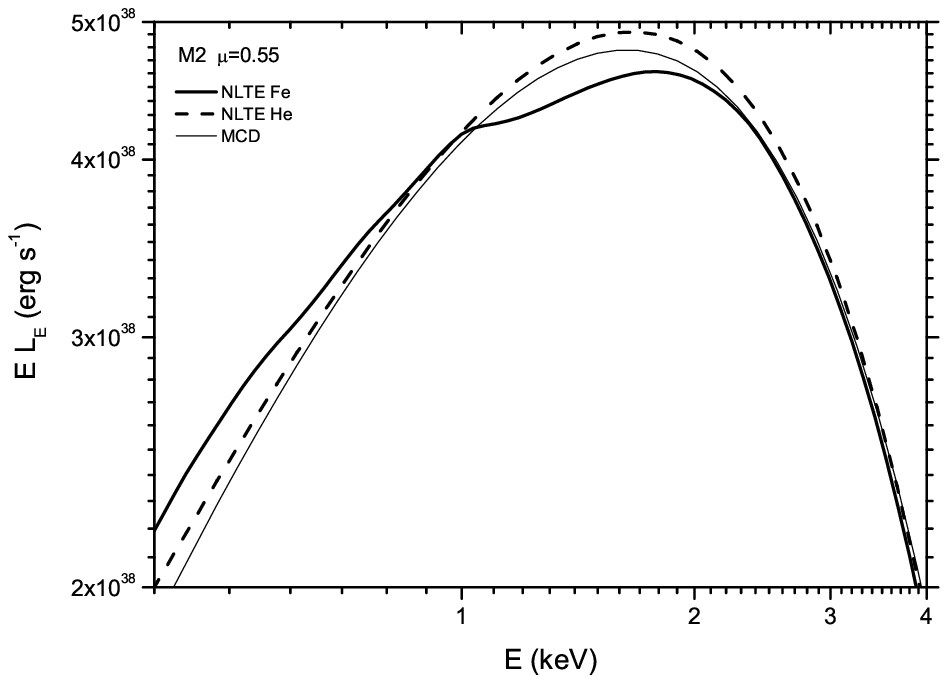}
\setlength{\abovecaptionskip}{0pt}
\setlength{\belowcaptionskip}{10pt} \caption{Spectra of accretion
disks with different chemical compositions in the observer's frame
at infinity. The central BH mass is $10^2~M_{\odot}$ and $a/M = 0.998$.
Similar legends are used as in
Fig.~\ref{fig:face-on}. The viewing angle, however, changes to
$\mu=0.55$.}\label{fig:oblique}
\end{figure}

Only in the extreme edge-on view does the MCD approximation
reproduce reasonably well the spectrum at energies above the peak.
When the observer is very close to the orbital plane, the whole
spectrum is shifted to much higher energies (the peak moves
from $E=0.81$~keV to $E=3.3$~keV, a factor of 4.1
shift), and the atomic fingerprints are totally eliminated by the
strong Doppler effect. Even so, the low-frequency deviations (now
all frequencies below 1.5~keV) persist unchanged.

\subsection{Limb-darkening law of integrated luminosity}

We have already discussed (\S~\ref{subsec:limb}) the angular
dependence of the emergent intensity, both in the fluid frame and
as seen by a distant observer, at selected energies.  Here
we explore the angle-dependence of the frequency-integrated
intensity as measured in the fluid frame and compare it
to approximations in the literature, such as the one used by
\citealt{gierlinski2001}. The results are displayed in
Fig.~\ref{fig:LDcheckM2}, which shows $r=1.6r_g$ in the
$100M_{\odot}$ extreme Kerr model with heavy elements.

Gierlinski et al. assumed the simple linear approximation
$I(\mu) \propto 1 + B\mu$ with $B=2.06$ everywhere.  On the
whole, its description of our full atmosphere
calculation is not too bad, both at this radius and at all others.
The frequency-integrated limb-darkening is in fact not far from
linear, and its slope ($B = 2.60\pm0.01$) is a bit greater
than the Gierlinski et al. assumed value.

However, there is still a deviation of as much as 15\% between
the full calculation and the linear approximation.  Virtually
all of this deviation can be eliminated by use of a different
analytic form with four parameters:
\begin{equation}\label{eq:newlimbdark}
I(\mu) \propto A_2+\frac{A_1-A_2}{1+e^{\frac{\mu-\mu_0}{\delta\mu}}}
\end{equation}
Choosing these as $A_1 = 0.108$, $A_2 = 0.44$, $\mu_0 = 0.48$,
and $\delta\mu = 0.2$ gives a remarkably good fit to the limb-darkening
behavior at all radii for both $100M_{\odot}$ and $1000M_{\odot}$
(Fig.~\ref{fig:LDcheckM2}).

\begin{figure}[ht]
\centering
\includegraphics[width=1.0\textwidth]{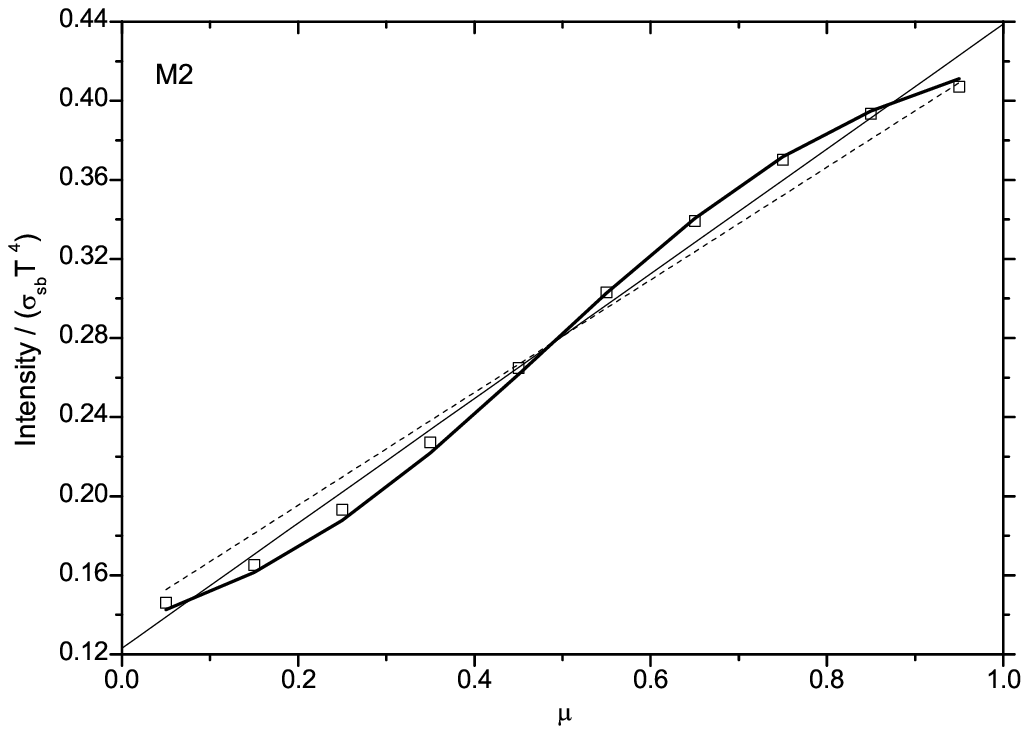}
\caption{Comparison of the limb-darkening law calculated by our
model to the linear law assumed by \citealt{gierlinski2001} in the
case of $10^2~M_{\odot}$.  The square data points are calculated
from our model at $r=1.6r_g$. The solid thick curve is generated using
eq.~\ref{eq:newlimbdark} and parameters discussed in the text.  The
limb-darkening law assumed by \citealt{gierlinski2001} is the
thin dashed line; our best-fit linear law is the thin solid
line.} \label{fig:LDcheckM2}
\end{figure}

\subsection{Observational Consequences}

   Although our models are far too few to span the likely range
of parameters applicable to IMBHs, their results have a number of
implications of immediate interest to the interpretation of
observational data.

\subsubsection{Temperature--Mass Ambiguities}

   The first observational implication is one that, at the qualitative
level, did not require calculations anywhere near as detailed as
ours to make.  We merely develop this point in a specific and
quantitative way.  It is that the mapping between characteristic
temperature of an accretion disk spectrum and the mass of the
black hole at its center is by no means 1:1.  As
Figure~\ref{fig:mass100} shows, when the spin parameter is large,
the peak energy of the spectrum can increase by as much as a
factor of 7 relative to what would be expected from a
Schwarzschild black hole. Part of this rise in temperature (for
fixed central mass and luminosity) is due to the ability of the
disk to extend closer to the black hole in the Kerr than in the
Schwarzschild metric; another part is due to Doppler shifting when
the viewing angle swings from pole-on to edge-on.  Because the
expected temperature scaling is $T \propto M^{-1/2}L^{1/4}$, the
inferred mass (assuming a Schwarzschild metric) depends on
observables as $M_{\rm inferred} \propto T^{-2} L^{1/2}$.  The
assumption of a non-spinning black hole could thus lead to an
underestimate of the central mass by as much as a factor of $\sim
50$.

    Conversely, as shown in Figure~\ref{fig:mass100_1000}, disks around
central masses differing by a factor of 10 can span a largely overlapping
range of temperatures if their parent black holes spin rapidly.  A larger
Doppler shift applied to light from a disk around a more massive black
hole can compensate for its intrinsically lower temperature.

     Thus, the characteristic energy--black hole mass relation is
ambiguous in both directions due to the effects of black hole spin.
Observers at different viewing angles can see a wide range
of characteristic energies in the light from a single black hole,
while a wide range of masses can all produce spectra with the same
characteristic energy.

\subsubsection{Angle-dependence in Bolometric Luminosity Estimates}

     Estimating the correct bolometric luminosity is subject to analogous
problems.  Reference to Figure~\ref{fig:mass100} also shows that
the {\it apparent} luminosity can vary by more than an order of
magnitude due solely to radiation anisotropy.  Unlike the
temperature shift, this problem is most severe when the black hole
rotates slowly if one has sufficiently broad-band spectral data.
However, with limited band-pass, the temperature shifts associated
with viewing angle changes in the maximal Kerr case can lead to
errors in the bolometric luminosity that are at least as large.

\subsubsection{Possible Visibility of Atomic Features}

     As we have stressed, one of our most important new results
is the influence of atomic photoionization features, principally
the K-edges of CVI and OVIII, on the emergent spectrum.  Best seen
in the spectra from disks around slowly-spinning black holes at
all viewing angles or in the spectra of disks around
rapidly-spinning black holes viewed face-on, these features create
several sharp breaks in slope near the peak of the spectrum when
$M \sim 100$--$1000M_{\odot}$, most prominently near 490~eV and
810~eV.  Because these features are due to H-like species, it is
possible that photoionization by hard X-rays could reduce their
strength; the extent to which that may be so depends on how much
of the hard X-ray continuum we see also shines on the annuli
principally responsible for these features, $10 \leq r/r_g \leq 20$
for OVIII, $20 \leq r/r_g \leq 30$ for CVI, for the parameters
considered here.   It is possible that the hard X-ray source is
sufficiently centrally concentrated or beamed outward as
to not greatly affect the feature-producing radii.  Detection of
(or bounds on) an Fe K$\alpha$ feature in these objects would
help constrain the amount and distribution of disk irradiation.

\section{Summary and Conclusion}\label{sec:conclusion}

   By constructing detailed atmosphere models for the disks around
intermediate mass black holes, we have shown that their spectra
possess a characteristic signature (almost) unique to the mass
range 100--$1000M_{\odot}$: atomic photoionization features near
the peak of the thermal spectrum.  These are present because
the energies of the K-edges in astrophysically abundant
medium-$Z$ elements are in the range 0.3--1~keV (depending on
ionization state and $Z$), and only in this mass range is
the (fluid frame) peak energy in the spectrum
of moderately sub-Eddington disks around black holes $\lesssim 1$~kev.
Higher mass black holes have lower
temperatures, so equivalent features due to H and He can appear
in disk spectra for $\sim 10^6$--$10^7M_{\odot}$ black holes, but
are also often subject to Comptonization smearing (\citealt{hubeny4}).
Lower mass black holes are so hot at their centers that all the
abundant elements but Fe are thoroughly stripped.

    A comparison of models with different central masses and
different spin parameters shows that there are sizable ambiguities
that interfere with linking observed parameters of X-ray sources
with the mass of the black hole responsible.  It is important
for the interpretation of observed spectra that these systematic
uncertainties be acknowledged.

    Finally, we have made a close comparison of the popular MCD
approximation with our more thorough calculations.  This comparison
is made somewhat multivalent by the numerous variations on this
approximation found in the literature (relativistic disk model or
simple $T \propto r^{-3/4}$?  which limb-darkening law?  relativistic
transfer function applied?).  To serve as our standard of comparison,
we have chosen the most careful of the various versions, that of
\citealt{gierlinski2001}.  We find that if the accretion
fuel came from Population II (or III) sources, the MCD approximation
would be fairly good, although there is a potentially observable
slope discrepancy at low energies.  However, the importance of
heavy element photoionization opacity in solar abundance material
at the temperatures prevalent in disks around black holes of
100--$1000M_{\odot}$ renders the MCD approximation much less
accurate.  Similarly, although the linear limb-darkening law
chosen by \citealt{gierlinski2001} gives a fairly good description
of the frequency-integrated angle-dependence of the emergent
intensity, it fails in interesting ways when applied to the
angle-dependence of the frequency-dependent intensity, particularly
for energies $> 1$~keV.

\clearpage

\acknowledgments

\end{document}